\documentclass[conference]{IEEEtran}
\usepackage{cite}
\usepackage{amsmath,amssymb,amsfonts}
\usepackage{graphicx}
\usepackage{textcomp}
\usepackage{url}
\usepackage{etoolbox}
\usepackage{cuted}
\usepackage{lipsum}
\usepackage{array}
\usepackage{color}
\usepackage{dirtytalk}
\usepackage{algpseudocode}
\usepackage{algorithm}

\usepackage[english]{babel}
\usepackage[utf8x]{inputenc}
\usepackage{amsmath,amsthm}

\usepackage{comment}
\usepackage{caption}

\theoremstyle{definition}

\theoremstyle{plain}
\newtheorem{theorem}{Theorem}

\ifCLASSINFOpdf
\else
\fi
\begin{document}
\title{Blockchain-enabled Identity Verification for Safe Ridesharing Leveraging Zero-Knowledge Proof}

\author{\IEEEauthorblockN{Wanxin Li\IEEEauthorrefmark{1}~~~~Collin Meese\IEEEauthorrefmark{1}~~~~Hao Guo\IEEEauthorrefmark{2}~~~~Mark Nejad\IEEEauthorrefmark{1}}
\IEEEauthorblockA{\IEEEauthorrefmark{1}
Department of Civil and Environmental Engineering, University of Delaware, U.S.A.\\\IEEEauthorrefmark{2}School of Software, Northwestern Polytechnical University, Taicang Campus, China.
 \\
\
\{wanxinli,cmeese,nejad\}@udel.edu \& haoguo@nwpu.edu.cn}}

\maketitle

\begin{abstract}
The on-demand mobility market, including ridesharing, is becoming increasingly important with e-hailing fares growing at a rate of approximately 130\% per annum since 2013. By increasing utilization of existing vehicles and empty seats, ridesharing can provide many benefits including reduced traffic congestion and environmental impact from vehicle usage and production. However, the safety of riders and drivers has become of paramount concern and a method for privacy-preserving identity verification between untrusted parties is essential for protecting users. To this end, we propose a novel privacy-preserving identity verification system, extending zero-knowledge proof (ZKP) and blockchain for use in ridesharing applications. We design a permissioned blockchain network to perform the ZKP verification of a driver's identity, which also acts as an immutable ledger to store ride logs and ZKP records. For the ZKP module, we design a protocol to facilitate user verification without requiring the exchange of any private information. We prototype the proposed system on the Hyperledger Fabric platform, with the Hyperledger Ursa cryptography library, and conduct extensive experimentation. To measure the prototype's performance, we utilize the Hyperledger Caliper benchmark tool to perform extensive analysis and the results show that our system is suitable for use in real-world ridesharing applications. 
\end{abstract}

\begin{IEEEkeywords}
Blockchain, data privacy, identity verification, ridesharing, zero-knowledge proof.
\end{IEEEkeywords}

\IEEEpeerreviewmaketitle

\section{Introduction}%Collin
Ridesharing services have received much attention in the past decade as an efficient method for increasing societal access to mobility using existing infrastructure. E-hailing fares have experienced significant growth at a rate of approximately 130\% per annum since 2013\cite{mckinseyrs}. In addition, a recent parking study shows that our vehicles are not in use for an average of 95\% of their lifetimes\cite{morris2016today}, presenting much room for increasing existing vehicle utilization. However, both regulators and users have expressed significant safety concerns with existing ridesharing applications, citing incidents such as driver impersonations, sexual assault and kidnapping. According to a recent report disclosed by Uber, there were nine assault-related deaths as well as over 5,900 cases of non-consensual sexual assault related incidents on their ridesharing platform in 2017 and 2018 combined\cite{ubersafety}. As a result, safeguarding the well-being of ridesharing users is a pivotal issue in the on-demand mobility market. 

Although some companies have implemented systems to help prevent misidentification and ensure safety, such as the Uber PIN system \cite{uberpin}, these schemes are centralized and platform exclusive, making them unsuitable for ensuring safety across the entire ridesharing ecosystem. Additionally, it is challenging to provide an extensive identity verification system while also respecting the privacy of untrusted users. Therefore, the problem of safe and privacy-preserving identity verification in ridesharing systems is of paramount interest and a solution is needed to enable universal and decentralized identity verification.

%Therefore, providing a universal and decentralized method for privacy-preserving user identification for use in ridesharing applications is of paramount interest if widespread societal adoption is to be achieved.

Recently, blockchain technology has been proposed as a plausible way of enabling efficient and decentralized two-sided sharing economies, such as ridesharing systems\cite{chang2018application}\cite{kato2018blockchain}\cite{8946128}. First implemented in Bitcoin: A Peer-to-Peer Electronic Cash System, %proposed by Satoshi Nakamoto in 2008,
blockchain technology is an emerging network technology enabling consensus among networked peers on a distributed, immutable digital ledger \cite{nakamoto2008bitcoin}. The inherent provenance and immutability properties of blockchain make it an ideal candidate for implementing a secure and decentralized identity verification protocol for future ridesharing systems. 

However, public blockchain systems present significant privacy concerns when sensitive user information is involved. With their transparency-by-design properties, any networked participant can view the entire contents of the ledger in a permissionless blockchain system such as Bitcoin. On the other hand, %a permissioned blockchain system (e.g., Hyperledger Fabric\cite{hyperledgerfabric}) allows for programmable access controls, making it suitable for constructing an identity verification protocol with properties of data ownership and privacy-preservation. Nonetheless, programmable access control serves only to protect the sensitive information stored on the ledger, but does not provide a method for preserving privacy during the verification process. 
zero-knowledge proof (ZKP) is a cryptographic scheme enabling one party to prove to another that they know a secret message without revealing any information other than possession of the secret. As a result, ZKP can be implemented as a module atop blockchain network as a way to protect the privacy of users during the verification process. 

In this paper, we propose a permissioned blockchain and zero-knowledge proof \cite{rackoff1991non} inspired approach for safe and privacy-preserving digital identity verification in existing ridesharing systems. This paper makes the following contributions:

\begin{itemize}
    \item We propose a novel zero-knowledge proof module atop a permissioned blockchain network for identity verification to protect the safety and privacy of riders and drivers without disclosing any sensitive information.
    
    %\item We design programmable access control policies for protecting trip records on the ledger, enabling users to define what entities can access their data and recording all retrieval events in an immutable access log.
    %To further increase the safety of ridesharing users, our scheme can be extended to also provide a means of undoubtedly verifying the driver's vehicle, for example by displaying the blockchain transaction number on a programmable LED sign. 
    %As a result, our proposed system enables a rider to both safely and undoubtedly verify that their driver is the correct one without exposing any of the driver's private information.
    
    \item We prototype the proposed system and perform benchmark tests on Hyperledger platform \cite{hyperledgerfabric}. The experimental results show that  our system is feasible for real-world ridesharing applications.

\end{itemize}
 
%\subsection{Organization}
The rest of this paper is organized as follows. In Section II, we present the requisite background knowledge regarding zero-knowledge proofs.
%and the Boneh-Lynn-Shacham (BLS) \cite{boneh2001short} signature scheme. 
A detailed overview of the proposed system architecture is described in Section III. To further illustrate the feasibility of our proposed scheme, we perform extensive experiments testing the performance of our identity verification system in Section IV. Also, we discuss the resilience of our proposed system against potential attacks in this section. For Section V, we present previous and related research in blockchain applications and blockchain-based ridesharing systems. Finally, we conclude this study in Section VI.

\section{Background Knowledge of Zero-Knowledge Proof}

%\subsection{Zero-Knowledge Proof}
Zero-knowledge proof (ZKP) was proposed in 1989 by Goldwasser, Micali, and Rackoff~\cite{goldwasser1989knowledge}.
In cryptography, a ZKP protocol is a method by which a prover can convince a verifier that he/she knows a secret message $m$, without conveying any information, apart from the fact that the prover knows the secret message $m$~\cite{wiki:zkp}. 
%The essence of zero-knowledge proofs is that, while it is trivial to prove one possesses knowledge of secret information by simply revealing the secret, it becomes challenging to prove such possession without disclosing the secret itself or any additional information. 
A ZKP of knowledge is a special case when the statement consists only of the fact that the prover possesses the secret information~\cite{wiki:zkp}. Based on the frequency of communications between the prover and the verifier, there are two kinds of ZKP schemes: Interactive ZKP and Non-interactive ZKP scheme. 
A zero-knowledge proof must satisfy three properties:

\begin{itemize}
    \item Completeness: If the statement is true, the honest verifier (that is, one following the protocol properly) will be convinced of this fact by an honest prover.
    \item Soundness: There is no such prover can convince an honest verifier if he/she does not compute the results correctly.
    %if the statement is false, no cheating prover can convince the honest verifier that it is true, except with some minuscule probability.
    \item Zero-knowledge: The proof of knowledge can be simulated without revealing any secret information which means that no verifier learns anything other than the fact that the statement is true.
    %if the statement is true, no verifier learns anything other than the fact that the statement is true. In other words, just knowing the statement (not the secret) is sufficient to imagine a scenario showing that the prover knows the secret. This is formalized by showing that every verifier has some simulator that, given only the statement to be proved (and no access to the prover), can produce a transcript that ``looks like" an interaction between the honest prover and the verifier in question.
\end{itemize}

ZKP can be applied to blockchain systems to address privacy issues. To be more specific, ZKP can be used to guarantee that transactions in blockchain are valid without revealing any sensitive information about the sender, the recipient or other transaction details. For instance, Zcash utilizes zk-SNARKs (zero-knowledge succinct non-interactive arguments of knowledge) scheme to allow blockchain peers to reach agreement on the validity of transactions without knowing details about the sender, recipient or amounts being transferred \cite{hopwood2016zcash}.

\section{System Architecture}%Wanxin

\begin{figure}[t]
\centering
\includegraphics[width=0.49\textwidth]{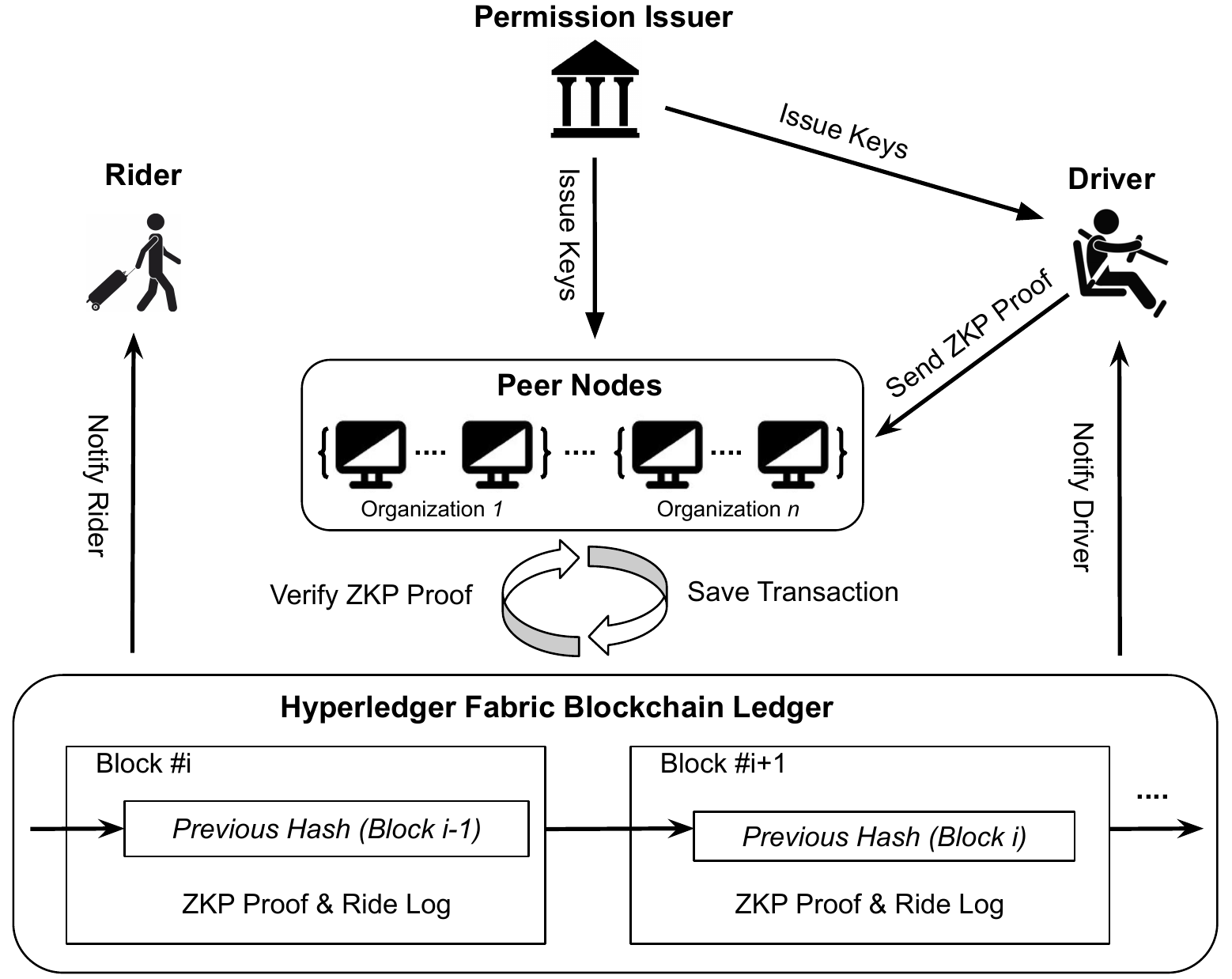}
\caption{Proposed System Architecture based on permissioned blockchain.}
\label{fig:arch}
\end{figure}

In this section, we describe the proposed privacy-preserving system architecture for identity verification in ridesharing, which includes a new zero-knowledge proof protocol and permissioned blockchain network. By referring to Fig. \ref{fig:arch}, we first define the following entities that take part in the proposed architecture:
\begin{itemize}
    \item Permission Issuer: A permission issuer is a trusted entity who issues identity information (e.g., driver license number) and key pairs to data owners and data verifiers. In practice, an agency such as Department of Motor Vehicles (DMV) can function as the permission issuer in the blockchain network.
    
    \item Driver: A driver is a client in blockchain network who owns his/her identity information and wants to prove their identity to riders without revealing the information.   
    
    \item Rider: A rider is a client in blockchain network who wants to safely verify the identity of driver before trip starts.
    
    \item Peer Node: A peer node is the element that hosts the ledger and smart contracts in the permissioned blockchain network and also acts as the data verifier who validates drivers' identities for riders. In our system, peer nodes are deployed and managed by a consortium of multiple organizations creating a decentralized network. %of peer nodes. %There are many devices that could function as peer nodes in our system, for example an existing roadside unit (RSU). 
    
    \item Permissioned Blockchain: A permissioned blockchain (in our prototype, Hyperledger Fabric) is utilized as the controller of the architecture and serves as the tamper-proof transaction ledger for saving proof records and trip information.

\end{itemize}

\begin{figure*}[ht]
\includegraphics[width=0.8\textwidth]{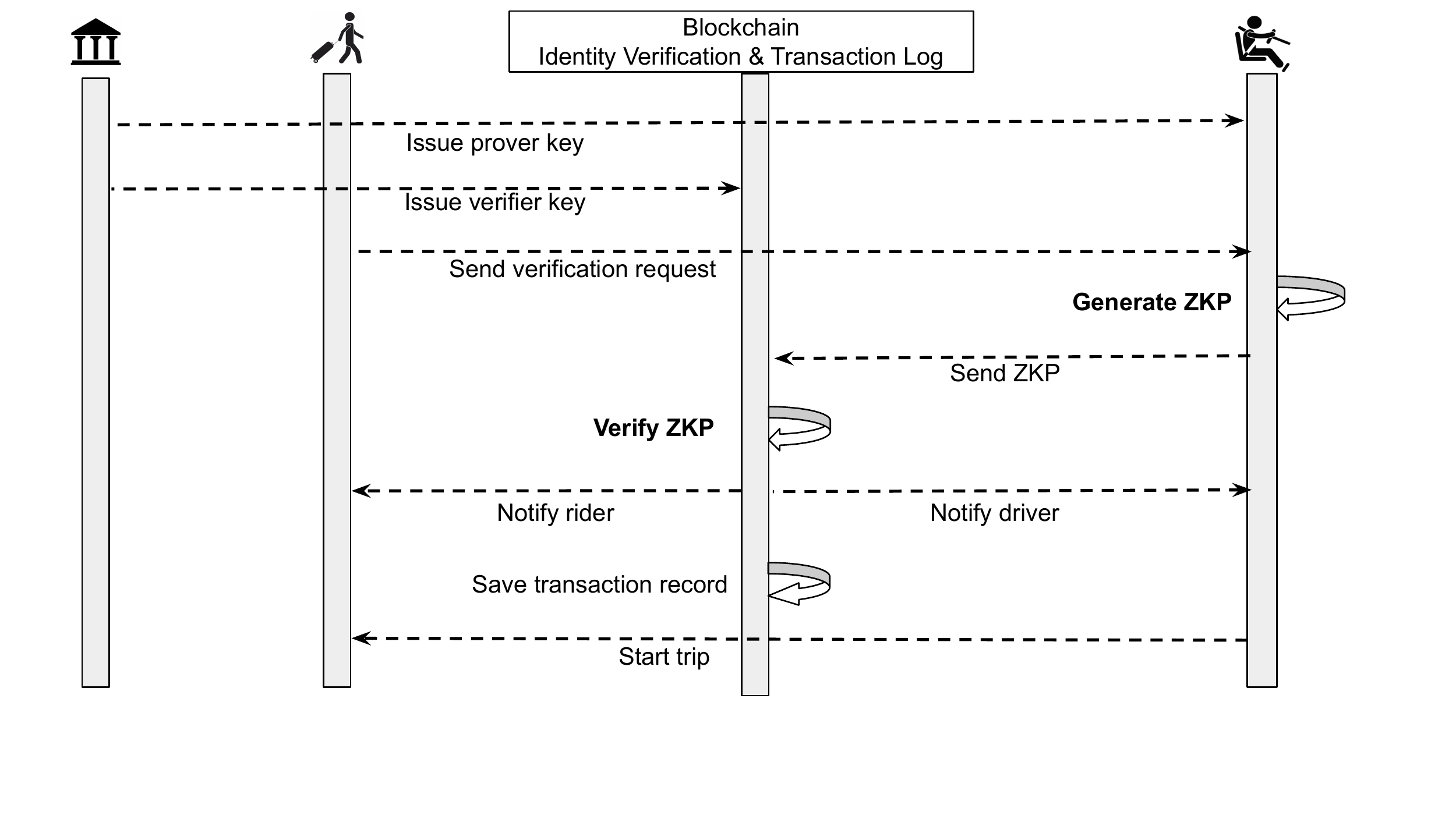}
\centering
\caption{Workflow of privacy-preserving identity verification in ridesharing scenario.}
\label{fig:workflow}
\end{figure*}

We design and develop the blockchain-based identity verification system for ridesharing on Hyperledger Fabric platform with Hyperledger Ursa cryptographic library \cite{hyperledgerursa}. Before each trip starts, the identity of a driver will be validated by the proposed zero-knowledge proof protocol without revealing the sensitive information to other participants in the network. The blockchain network maintains a distributed ledger for recording trip information and proof records. In the remainder of this section, we illustrate the design of zero-knowledge proof protocol, followed by a description of the workflow of identity verification in ridesharing using our proposed system. We discuss the implementation of the blockchain network in Section \ref{blockchain module}.

\subsection{Zero-Knowledge Proof Protocol}
As shown in Fig. \ref{fig:arch}, we introduce the zero-knowledge proof protocol for verifying the identity of drivers without revealing sensitive information to other participants. When a ridesharing service matches a driver with a rider, the driver acts as a prover to prove his/her identity to a peer node from blockchain network, acting as a verifier, in ZKP-based encrypted message. After the blockchain peer validates the driver's identity, the result is communicated to both parties. 

%The traversing vehicle acts as a prover to prove its identity to a gateway, acting as a verifier, in a ZKRP-based encrypted message.

%After a ridesharing service matches the rider and driver, the driver acts as prover to prove his/her identity to a peer node from blockchain network, acting as a verifier, in ZKP-based encrypted message.

\begin{theorem}
\label{theorem-1}
Let $G$ be a multiplicative cyclic group of prime order $p$ with generator $g$. Let $e : G \times G \rightarrow G_T$ be a computable, bilinear and non-degenerate pairing into the group $G_T$. Then, we have $e(x, y) = e(y, x)$ for all $x,y \in G$ because $G$ is cyclic.
\end{theorem}

Based on Theorem \ref{theorem-1} \cite{cyclic}, we describe how to construct the ZKP protocol in our proposed scheme for identity verification. In addition to the following, we detail the specifics of our proposed ZKP scheme in the context of ridesharing in Section \ref{zkp-module}. In the prototype, we choose BLS scheme \cite{boneh2001short} to build the generator $g$ and elliptic curve \cite{frey1999tate} for bilinear pairing $e$. Our ZKP protocol consists of three key functions:

\subsubsection{Key Generation}
The key generation algorithm selects a random $a \in \mathbb{Z}_p$ and computes $v = g^a$. The prover key is $a \in \mathbb{Z}_p$ and the verifier key is $v \in G$.

\subsubsection{Proof Generation}
Given the prover key $a$, and the identity information $m$, the prover computes the hashed $m$ in SHA256 algorithm \cite{rachmawati2018comparative}, as $h = H(m)$. Then, the proof is generated as $\delta = h^a \in G$.

\subsubsection{Proof Verification}
Given the proof $\delta$ and the verifier key $v$, the verifier can verify that $e(\delta, g) = e(h, v) = e(H(m), g^a)$ without revealing the identity information $m$, reject otherwise.

\subsection{Identity Verification Workflow in Ridesharing}
The workflow of identity verification for ridesharing is depicted in Fig. ~\ref{fig:workflow}. In the beginning, all the drivers and riders register themselves in the blockchain network. The permission issuer (e.g., Department of Motor Vehicles) issues prover keys to drivers and verifier keys to peer nodes. The registration and key issuance process only needs to be performed once for each user. When a ridesharing service is matched between a driver and a rider, the rider first sends the verification request to the driver prior to trip start. After receiving the request from rider, the driver uses the prover key to generate zero-knowledge proof for his identity information and send it to a peer node in blockchain network. Next, the peer node uses the verifier key to validate the zero-knowledge proof from the driver. When the verification is complete, the blockchain network will notify both the rider and the driver of the result. Simultaneously, a smart contract is executed to record this transaction with trip details on the ledger. After that, the driver can start the trip. %The specifics of the dynamic matching and pricing systems are outside the scope of this paper but will be explored in future works. 

% \begin{figure*}[ht]
% \includegraphics[width=0.8\textwidth]{Figures/workflow0707.pdf}
% \centering
% \caption{Workflow of privacy-preserving identity management in ridesharing scenario.}
% \label{fig:workflow}
% \end{figure*}

\section{Experiments and Evaluation}%Wanxin
\subsection{Experimental Setup}
We prototype the proposed identity verification system and conduct a series of experiments to evaluate its performance. The system consists of two primary portions that interact seamlessly: the ZKP module and the blockchain network. The ZKP module is programmed by using the Hyperledger Ursa library \cite{hyperledgerursa}. The blockchain network is developed on the Hyperledger Fabric v1.2 and tested using Hyperledger Caliper benchmark tool \cite{hyperledgercaliper}. For testing, we instantiate 10 clients including 5 drivers and 5 riders, in the blockchain network. The prototype and experiments are deployed and conducted on multiple Fabric peers in Docker containers locally on Ubuntu 18.04 operating system with 2.8 GHz Intel i5-8400 processor and 8GB DDR4 memory.

\subsection{ZKP Module}
\label{zkp-module}
As illustrated in Fig. \ref{fig:zkp-process}, the ZKP module performs the functionalities of initial setup, generation and verification of zero-knowledge proofs of drivers' identity information. These functionalities are programmed by using Hyperledger Ursa, a cryptographic library for Hyperledger applications. Hyperledger Ursa is programmed using the Rust language and provides APIs for various cryptographic schemes. Our ZKP module operates in the following three phases:

\begin{figure}[t]
\centering
\includegraphics[width=0.45\textwidth]{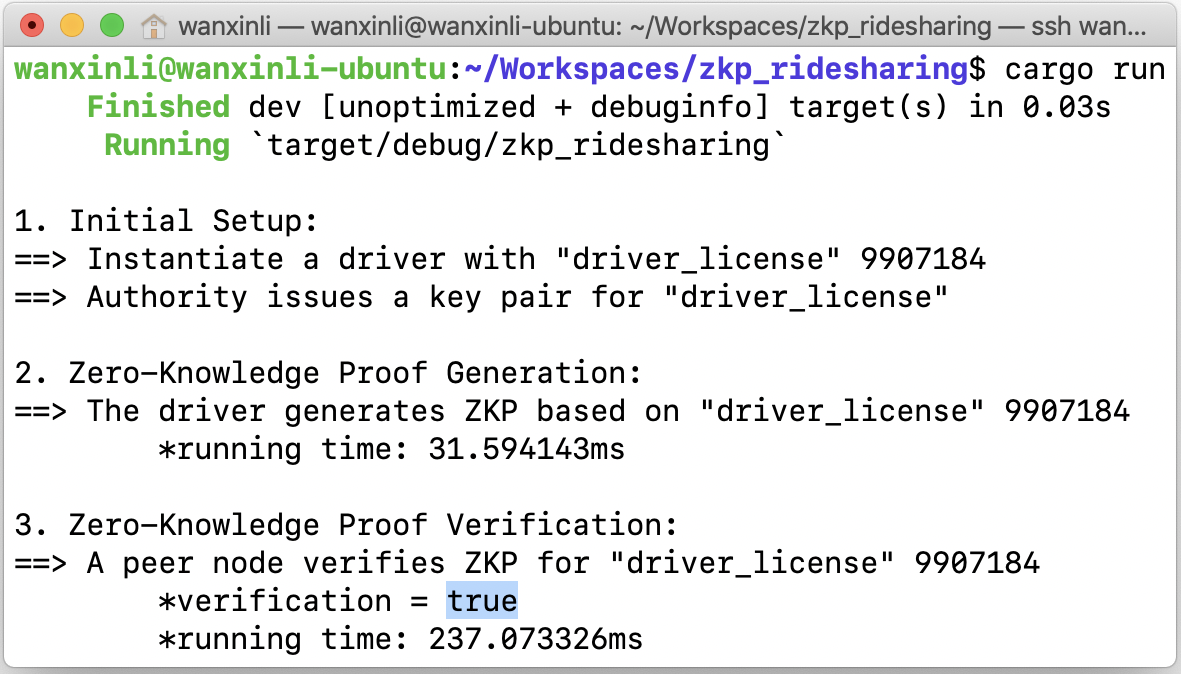}
\caption{Process of the ZKP module.}
\label{fig:zkp-process}
\end{figure}

\subsubsection{Phase 1 - Initial Setup}
Phase 1 initializes a driver instance to act as the prover. As shown in Fig. \ref{fig:zkp-process}, the driver has the identity information {\tt driver\_license} (value: {\tt 9907184}) issued by the Department of Motor Vehicles, and the permission issuer generates a key pair for the identity information it issues. The BLS scheme \cite{boneh2001short} is used to build the key pair generator, which creates the prover key for the driver and the verifier key for the peer node in blockchain network, as follows:

\begin{verbatim}
let generator = Generator::new().unwrap();
let prover_key = SignKey::new().unwrap();
let verifier_key = VerKey::new(&generator,
                 &sign_key).unwrap();
\end{verbatim}

%The BLS scheme uses a bilinear pairing for verification, and signatures are elements of an elliptic curve group.

\subsubsection{Phase 2 - ZKP Generation}
In this phase, the driver uses the prover key to generate a one-time zero-knowledge proof for the hashed {\tt driver\_license} via SHA256 algorithm \cite{rachmawati2018comparative}. The resulting proof consists of three elements on an elliptic curve. For instance, as shown below:

\begin{figure}[h]
\centering
\includegraphics[width=0.49\textwidth]{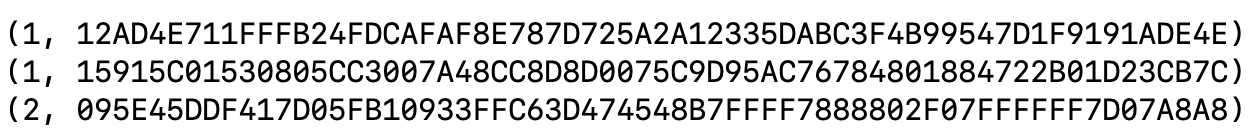}
%\caption{Example of generated zero-knowledge proof.}
\label{fig:proof-example}
\end{figure}

\noindent
the proof of hashed {\tt 9907184} from {\tt driver\_license} is a combination of three points on an elliptic curve represented in hexadecimal format. Our experiments show that the average running time for proof generation phase is 32 ms. 

%{\tiny(1, 12AD4E711FFFB24FDCAFAF8E787D725A2A12335DABC3F4B99547D1FD1F9191ADE4E)}

\subsubsection{Phase 3 - ZKP Verification}
After proof generation, a peer node from the blockchain network (Section \ref{blockchain module}) verifies the zero-knowledge proof from the driver. The verification function takes the proof, the hashed {\tt driver\_license}, the verifier key and the corresponding generator as inputs, and utilizes elliptic curve bilinear pairing \cite{frey1999tate} to verify the proof:

\begin{verbatim}
let result = Bls::verify($proof, 
             driver_license.as_slice(),
             $verifier_key, $generator)
             .unwrap();
\end{verbatim}

In our experiments, the average running time for verifying each zero-knowledge proof is around 239 ms. After that, the blockchain system can authenticate the driver's identity anonymously and subsequently communicate the result to both the rider and driver.

\begin{figure}[t]
\centering
\includegraphics[width=0.48\textwidth]{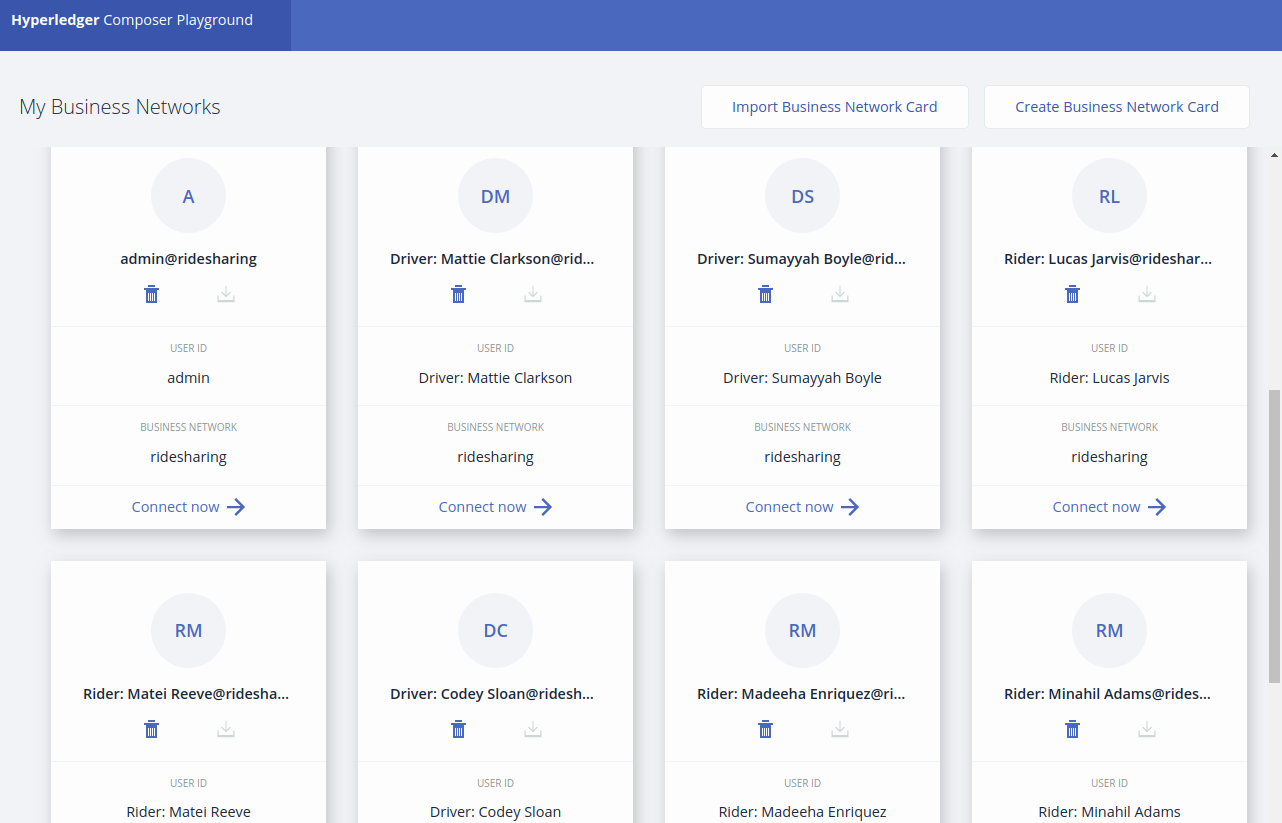}
\caption{Blockchain-based ridesharing network login window.}
\label{fig:login-window}
\end{figure}

\subsection{Blockchain Network}
\label{blockchain module}
Hyperledger Composer~\cite{hcomposer} is a framework to build and run applications on top of Hyperledger Fabric. %(Fig. \ref{fig:hyperledger-fabric}). 
Hyperledger Composer provides four programmable modules: Model File, Script File, Access Control and Query File. We first define all the objects in the Model File including the data structures of participants and transactions. Then, smart contracts are written, using Chaincode \cite{chaincode}, in the Script File. The access control policies are coded in the Access Control File to facilitate different access levels to different participants. As for the Query File, it works similar to conventional database query operations for data retrieval. Lastly, these files are packaged up into one Business Network Archive (.bna) file and deployed on the blockchain network. 

The proposed blockchain system also provides a web interface for participants interacting with the blockchain network as shown in the Fig. \ref{fig:login-window}. The traffic management authority (e.g., United States Department of Transportation) acts as the administrator to issue access permissions for the other clients including riders and drivers. As a result, each participant has an ID registry for connecting to the blockchain network.

\begin{figure}[t]
\centering
\includegraphics[width=0.45\textwidth]{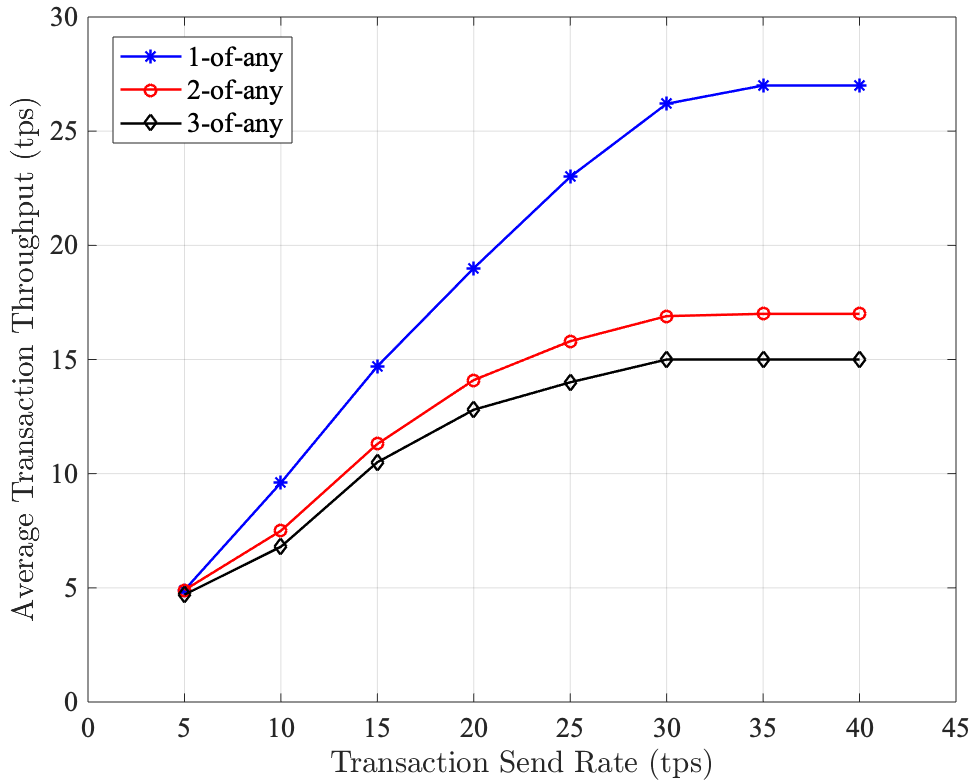}
\caption{Average transaction throughput vs. send rate under different Hyperledger Fabric endorsement policies.}
\label{fig:throughput-sendrate}
\end{figure}

\subsection{Performance Evaluation}
In this subsection, we evaluate the performance of both the ZKP module and blockchain network. To measure the performance of the ZKP module, we conduct experiments to analyze the effects of varying the length of the secret message. For blockchain network evaluation, we perform benchmark tests using Hyperledger Caliper with varying settings for transaction send rates and endorsement policies. The definitions of these two benchmark factors are as follows:

\begin{itemize}
    \item Transaction Send Rate: The transaction send rate defines the rate at which transactions are input to the blockchain network system, which is a key factor for stress testing.
    
    \item Endorsement Policy: The endorsement policies define the set of peers need to reach agreement on the result of a transaction before it can be committed to the ledger.
    
\end{itemize}

\subsubsection{ZKP Performance}
%To evaluate the performance of the ZKP module, we conduct experiments to analyze the effect of varying the length of the secret. 
The default length of the secret message ({\tt driver\_license}) is 7 characters using String type. In our experiments, we increase the length of secrets from 1 to 10, 100, 1,000 and 10,000 characters and then measure the results. The results %(Fig. \ref{fig:secret-length}) 
show that both proof generation and verification times remain constant regardless of the secret length, with times of approximately 32 ms and 239 ms, respectively. %explain the reason.
Our ZKP module can offer constant proof generation and verification time because a secret message $m$ is hashed to a fixed length of 256-bit value via SHA256 algorithm \cite{rachmawati2018comparative} before proof generation. %In the experiments, we invoke {\tt new\_proof\_builder} and {\tt new\_proof\_verifier} functions from the Ursa library, which utilize a HashMap to compute and verify the proof. 
As a result, the proof generation and verification time are independent from the length of secret messages.
Non-reliance on secret length provides our ZKP scheme with more flexibility for verifying different kinds of secret values (e.g., social security number, taxpayer identification number, etc.) without sacrificing security and efficiency. Compared to the signing phase, the verification phase takes additional time because it requires computing two pairings on the elliptic curve \cite{boneh2001short}.

% \begin{figure}[h]
% \centering
% \includegraphics[width=0.46\textwidth]{Figures/zkp0709.png}
% \caption{Run time of ZKP vs. the length of secrets.
% }
% \label{fig:secret-length}
% \end{figure}

\subsubsection{Transaction Throughput}
The transaction throughput measures the flow rate of processed transactions through the blockchain network, in units of transactions per second, during the test cycle. As shown in Fig. \ref{fig:throughput-sendrate}, when increasing the transaction send rate, the average transaction throughput will increase in the beginning and then reach peaks at 27 tps, 17 tps, and 15 tps under 1-of-any, 2-of-any, and 3-of-any endorsement policies, respectively. The choice of endorsement policy can impact the transaction throughput. For example, with a fixed transaction send rate of 20 tps, when increasing the number of endorsing peers, the average transaction throughput will decrease. This is due to the fact that more endorsing peers increase the complexity of the endorsement process.

\begin{comment}
In addition, we conduct multiple rounds of tests to record the minimum, average and maximum transaction throughputs. The results show that the difference between minimum and maximum transaction throughputs is relatively small under different endorsement policies. 

\begin{figure}[t]
\centering
\includegraphics[width=0.4\textwidth]{Figures/throughput_endorse-0713.png}
\caption{Minimum, average and maximum transaction throughput vs. Hyperledger Fabric endorsement policy under the send rate of 20 tps.}
\label{fig:throughput-endorse}
\end{figure}
\end{comment}

% \begin{figure}[h]
% \centering
% \includegraphics[width=0.4\textwidth]{Figures/throughput_endorse-0712.png}
% \caption{Average transaction throughput vs. Hyperledger Fabric endorsement policies under different send rates.}
% \label{fig:throughput-endorse}
% \end{figure}

\subsubsection{Transaction Latency}
\begin{figure}[t]
\centering
\includegraphics[width=0.45\textwidth]{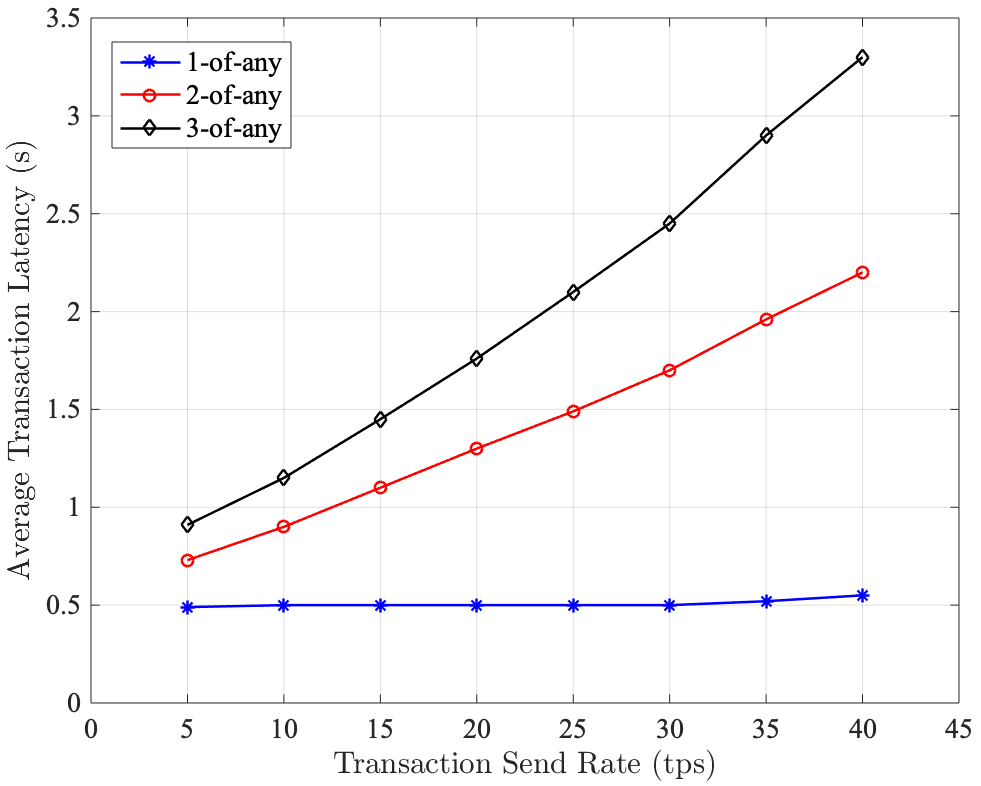}
\caption{Average transaction latency vs. send rate under different Hyperledger Fabric endorsement policies.}
\label{fig:latency-sendrate}
\end{figure}

The transaction latency indicates the processing time of a transaction, from submission by the client until it is processed and committed to the ledger. As shown in Fig. \ref{fig:latency-sendrate}, when increasing the transaction send rate, the average transaction latency will also increase significantly for 2-of-any and 3-of-any endorsement policies. However, the average transaction latency remains constant at 0.5 second under the 1-of-any endorsement policy, as long as the transaction send rate is lower than 30 tps. When increasing the transaction send rate beyond 30 tps, the average latency begins to increase, but at a significantly slower rate in comparison with the of the 2-of-any or 3-of-any policies. This is because the 1-of-any endorsement policy has a higher transaction throughput threshold due to its reduced complexity. In addition, the choice of endorsement policy can significantly impact the transaction latency. For example, with a fixed transaction send rate of 20 tps, when increasing the number of endorsement peers, the average transaction latency will increase. 

\begin{comment}

Moreover, we conduct multiple rounds of tests to record the minimum, average and maximum transaction latencies. The results show that the difference between minimum and maximum transaction latencies will also increase when increasing the number of endorsing peers.

% \begin{figure}[h]
% \centering
% \includegraphics[width=0.4\textwidth]{Figures/latency_endorse-0712.png}
% \caption{Average transaction latency vs. Hyperledger Fabric endorsement policies under different send rates.}
% \label{fig:latency-endorse}
% \end{figure}

\begin{figure}[t]
\centering
\includegraphics[width=0.4\textwidth]{Figures/latency_endorse-0713.png}
\caption{Minimum, average and maximum transaction latency vs. Hyperledger Fabric endorsement policy under the send rate of 20 tps.}
\label{fig:latency-endorse}
\end{figure}

\end{comment}

\subsubsection{Success Rate}
The success rate measures how many transactions, out of all submitted transactions, have been successfully processed and written on the blockchain during a test cycle. A failed transaction could be due to network time-outs, incorrect network configuration or bugs in smart contract. For all test cycles, regardless of different send rates and endorsement policies, our blockchain network can always achieve 100\% success rates.

\subsection{Discussion on Potential Attacks}
In this subsection, we discuss the resilience of the proposed privacy-preserving identity verification system against three potential attacks.

\subsubsection{Ride Log Tampering Attack}
In the proposed system, a newly generated ride log is stored on the blockchain in the form of a transaction. 
%Meanwhile, the permanently recorded ride log is also protected by access control policies.
%Meanwhile, the ride log is also recorded permanently on the blockchain network, which is protected by access control policies.
%When a malicious attacker attempts to tamper the existing ride log,  the newly generated hash digest will be different from the original hash digest stored in the blockchain. As a result, participants immediately find that the ride log has been attacked.
Our system will reject any attempt to tamper an existing ride log due to the immutable feature of blockchain, ensuring data integrity for all recorded information. Therefore, once the trip information is recorded, it is protected from modification by any entity. %We simulated the attacking strategy~\cite{chen2018exposing} by trying to tamper vehicular information on the ledger. 
%Thus, any attempt to tamper an existing record will be immediately rejected by our system.
%Moreover, the proposed scheme enforces identity-based access control policies on the ride log which will also protect the security of travelling data since an unauthorized user cannot obtain the access permission.

\subsubsection{Eavesdropping Attack}
Our proposed ZKP module can protect ridesharing systems from the eavesdropping attack. In this attack, the malicious attacker intercepts the message between driver (a prover) and peer node (a verifier) from the blockchain network in order to steal or gain access to confidential information. In the worst case, if the sensitive information has been accessed by the attacker, it still cannot reveal the actual information since the identity information is ZKP-encrypted. As a result, no sensitive information is disclosed even in the occurrence of an eavesdropping attack.

\subsubsection{Fake Driver Attack}
\begin{figure}[t]
\centering
\includegraphics[width=0.45\textwidth]{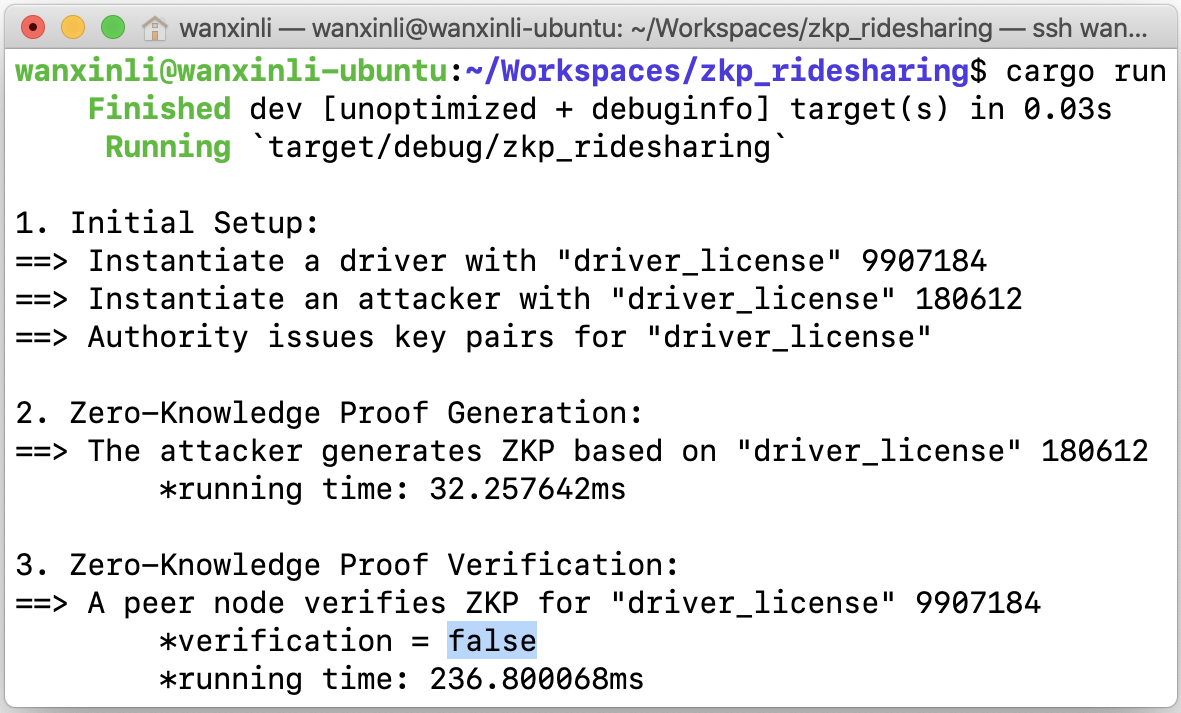}
\caption{Response against fake driver attack.}
\label{fig:zkp-attack}
\end{figure}

A fake driver attack is a situation where the driver wants to spoof the ZKP verification process by providing fake identity information to the verifier.
%in which an attacker belongs to one blockchain network $N(\epsilon)$, but it spoofs the gateway that it belongs to another blockchain network $N(\delta)$. 
%In this scenario, validating identity without revealing the sensitive information becomes a challenging tasks. 
However, the ZKP protocol is resistant to fake identity s
poofing by design, protecting against this type of attack.
%if the attacker successfully injects false login status among multiple blockchain networks.
For instance, if the attacker intends to deceive the prover, but actually has the incorrect identity (e.g., {\tt driver\_license 180612}), our system will detect and reject the malicious proof. For instance, the output of our protocol against a fake driver attack is shown in Fig. \ref{fig:zkp-attack}.

\section{Related Work}%Collin
In this section we outline the related work on both blockchain in general and blockchain-based ridesharing systems which motivate and compliment our research.

Blockchain distributed network technology has received much research attention since its inception and studies have demonstrated its applicability to a variety of areas including decentralized finance, healthcare, Internet of Things (IoT) and sharing economies. For example, in the area of decentralized finance, Desai et al. \cite{8946198} present a hybrid blockchain architecture for privacy-enabled and accountable auctions and \cite{8946175} details a blockchain-enabled croudsourcing platform that preserves user privacy. Additionally, a platform for initial public offerings on a permissioned blockchain system is outlined in \cite{8946171}. In the area of healthcare, \cite{8946149} proposes a novel method for secure access control in electronic health record management using a hybrid blockchain architecture and attribute-based multiple signatures \cite{9169395}. Regarding blockchain-based IoT systems, Kim et al. \cite{8946226} introduce SSP: self-sovereign privacy for IoT using a combination of blockchain and Multi-Party-Computations (MCP). Wanxin et al. \cite{li2019blockchain} introduce a blockchain-based approach for defending emerging Intelligent Traffic Signal Systems against malicious attackers and then extend the study in multiple vehicular networks \cite{li2020privacy}. Hao et al. propose a blockchain-inspired event recording system for autonomous vehicles \cite{9217515}.

There have been various attempts to address the problems of centralized ridesharing, and sharing economies in general, using blockchain-based systems. 
%The authors of \cite{aivodji2016meeting} proposed an approach to protect the privacy of location data of users in decentralized ridesharing services.
%In this design, local computations, secure MCP and multimodal routing algorithms are combined to generate a ridesharing itinerary for users while preserving their private geographical information. %Additionally, some studies have proposed an all-inclusive approach to decentralized ridesharing using public blockahin systems. 
Baza et al. \cite{baza2019b} propose a privacy-preserving and trustless ridesharing application with proof of concept implemented atop the Ethereum blockchain \cite{etherwhitepaper}. Their system includes protocols for matching, payment, reputation and identity management utilizing a combination of smart contracts and zero-knowledge-range-proofs. However, their system incorporates a fully decentralized public blockchain which can dramatically increases resource consumption and transaction confirmation time, as well as associated identity verification latency, in comparison to a permissioned blockchain design such as in our system. 

\cite{kanza2018cryptotransport} and \cite{baza2020light} also propose comprehensive ridesharing systems atop a public blockchain which preserves privacy using pseudonymity schemes. These designs motivated our research into a system for identity management and verification for ridesharing which requires no pseudonyms or exchange of private information between either party. Additionally, Semenko et al. \cite{semenko2019distributed} propose a decentralized and privacy-preserving ridesharing platform using direct two-party encryption and smart contract based access control. The result, however, is that users must exchange and reveal their sensitive information to other parties if they wish to use the service. %Lastly, in \cite{zhang2019blockchain} a distributed and privacy-preserving ridesharing system is proposed with a focus on the package delivery application. A hash-oriented Practical Byzantine Fault Tolerance (PBFT) algorithm is designed to reduce the time to achieve consensus from minutes to 15 seconds. That being said, their design exclusively uses mail box locations for pickups and dropoffs to avoid revealing the location of package owners as a privacy-preservation method, which is not applicable to our design. 

\section{Conclusion}
In this paper, we present a novel privacy-preserving identity verification system, extending zero-knowledge proof atop blockchain for use in ridesharing applications. Our proposed scheme enables secure verification without requiring exchange of any sensitive information between untrusted parties. %Access control policies are designed in the system to enable users to manage what entities can access their trip records on the ledger. 
We prototype the system and perform extensive experiments to assess its performance and practicality under varying conditions. Our results show the blockchain network, based on Hyperledger Fabric, provides high transaction throughput with low-latency. Meanwhile, the ZKP module is able to perform identity verification within milliseconds, making our design suitable for use in real-world ridesharing applications. In addition, we discuss the resilience of our system against multiple potential attacks. 

\Urlmuskip=0mu plus 1mu\relax

\bibliographystyle{IEEEtran}
\bibliography{sigproc.bib}

\begin{thebibliography}{10}
\providecommand{\url}[1]{#1}
\csname url@samestyle\endcsname
\providecommand{\newblock}{\relax}
\providecommand{\bibinfo}[2]{#2}
\providecommand{\BIBentrySTDinterwordspacing}{\spaceskip=0pt\relax}
\providecommand{\BIBentryALTinterwordstretchfactor}{4}
\providecommand{\BIBentryALTinterwordspacing}{\spaceskip=\fontdimen2\font plus
\BIBentryALTinterwordstretchfactor\fontdimen3\font minus
  \fontdimen4\font\relax}
\providecommand{\BIBforeignlanguage}[2]{{%
\expandafter\ifx\csname l@#1\endcsname\relax
\typeout{** WARNING: IEEEtran.bst: No hyphenation pattern has been}%
\typeout{** loaded for the language `#1'. Using the pattern for}%
\typeout{** the default language instead.}%
\else
\language=\csname l@#1\endcsname
\fi
#2}}
\providecommand{\BIBdecl}{\relax}
\BIBdecl

\bibitem{mckinseyrs}
\BIBentryALTinterwordspacing
T.~Baltic, A.~Cappy, R.~Hensley, and N.~Pfaff. How sharing the road is likely
  to transform american mobility. [Online]. Available:
  \url{https://www.mckinsey.com/industries/automotive-and-assembly/our-insights/how-sharing-the-road-is-likely-to-transform-american-mobility#}
\BIBentrySTDinterwordspacing

\bibitem{morris2016today}
D.~Z. Morris, ``Today’s cars are parked 95\% of the time,'' \emph{Fortune,
  March}, vol.~13, 2016.

\bibitem{ubersafety}
\BIBentryALTinterwordspacing
Uber. Uber's united safety report. [Online]. Available:
  \url{https://www.uber.com/us/en/about/reports/us-safety-report/}
\BIBentrySTDinterwordspacing

\bibitem{uberpin}
\BIBentryALTinterwordspacing
------. For drivers: How pin verification works. [Online]. Available:
  \url{https://www.uber.com/blog/pin-verification-drivers/}
\BIBentrySTDinterwordspacing

\bibitem{chang2018application}
S.~E. {Chang} and C.~{Chang}, ``Application of blockchain technology to smart
  city service: A case of ridesharing,'' in \emph{2018 IEEE International
  Conference on Internet of Things (iThings) and IEEE Green Computing and
  Communications (GreenCom) and IEEE Cyber, Physical and Social Computing
  (CPSCom) and IEEE Smart Data (SmartData)}, 2018, pp. 664--671.

\bibitem{kato2018blockchain}
K.~{Kato}, Y.~{Yan}, and H.~{Toyoizumi}, ``Blockchain application for rideshare
  service,'' in \emph{2018 8th International Conference on Logistics,
  Informatics and Service Sciences (LISS)}, 2018, pp. 1--5.

\bibitem{8946128}
P.~{Pal} and S.~{Ruj}, ``Blockv: A blockchain enabled peer-peer ride sharing
  service,'' in \emph{2019 IEEE International Conference on Blockchain
  (Blockchain)}, 2019, pp. 463--468.

\bibitem{nakamoto2008bitcoin}
S.~Nakamoto, ``Bitcoin: A peer-to-peer electronic cash system,'' 2008.

\bibitem{rackoff1991non}
C.~Rackoff and D.~R. Simon, ``Non-interactive zero-knowledge proof of knowledge
  and chosen ciphertext attack,'' in \emph{Annual International Cryptology
  Conference}.\hskip 1em plus 0.5em minus 0.4em\relax Springer, 1991, pp.
  433--444.

\bibitem{hyperledgerfabric}
\BIBentryALTinterwordspacing
{Hyperledger Fabric}. [Online]. Available:
  \url{https://www.hyperledger.org/projects/fabric}
\BIBentrySTDinterwordspacing

\bibitem{goldwasser1989knowledge}
S.~Goldwasser, S.~Micali, and C.~Rackoff, ``The knowledge complexity of
  interactive proof systems,'' \emph{SIAM Journal on computing}, vol.~18,
  no.~1, pp. 186--208, 1989.

\bibitem{wiki:zkp}
\BIBentryALTinterwordspacing
{Wikipedia contributors}, ``Zero-knowledge proof --- {Wikipedia}{,} the free
  encyclopedia,'' 2020. [Online]. Available:
  \url{https://en.wikipedia.org/wiki/Zero-knowledge_proof}
\BIBentrySTDinterwordspacing

\bibitem{hopwood2016zcash}
D.~Hopwood, S.~Bowe, T.~Hornby, and N.~Wilcox, ``Zcash protocol
  specification,'' \emph{GitHub: San Francisco, CA, USA}, 2016.

\bibitem{hyperledgerursa}
\BIBentryALTinterwordspacing
{Hyperledger Ursa}. [Online]. Available:
  \url{https://www.hyperledger.org/projects/ursa}
\BIBentrySTDinterwordspacing

\bibitem{cyclic}
\BIBentryALTinterwordspacing
{Cyclic Group Supplement}. [Online]. Available:
  \url{https://www.math.lsu.edu/~adkins/m4200/cyclicgroup.pdf}
\BIBentrySTDinterwordspacing

\bibitem{boneh2001short}
D.~Boneh, B.~Lynn, and H.~Shacham, ``Short signatures from the weil pairing,''
  in \emph{International conference on the theory and application of cryptology
  and information security}.\hskip 1em plus 0.5em minus 0.4em\relax Springer,
  2001, pp. 514--532.

\bibitem{frey1999tate}
G.~Frey, M.~Muller, and H.-G. Ruck, ``The tate pairing and the discrete
  logarithm applied to elliptic curve cryptosystems,'' \emph{IEEE Transactions
  on Information Theory}, vol.~45, no.~5, pp. 1717--1719, 1999.

\bibitem{rachmawati2018comparative}
D.~Rachmawati, J.~Tarigan, and A.~Ginting, ``A comparative study of message
  digest 5 (md5) and sha256 algorithm,'' in \emph{Journal of Physics:
  Conference Series}, vol. 978, no.~1, 2018, p. 012116.

\bibitem{hyperledgercaliper}
\BIBentryALTinterwordspacing
{Hyperledger Caliper}. [Online]. Available:
  \url{https://www.hyperledger.org/use/caliper}
\BIBentrySTDinterwordspacing

\bibitem{hcomposer}
\BIBentryALTinterwordspacing
{Hyperledger Composer}. [Online]. Available:
  \url{"https://www.hyperledger.org/projects/composer"}
\BIBentrySTDinterwordspacing

\bibitem{chaincode}
\BIBentryALTinterwordspacing
{Chaincode}. [Online]. Available:
  \url{https://hyperledger-fabric.readthedocs.io/en/release-1.4/chaincode.html}
\BIBentrySTDinterwordspacing

\bibitem{8946198}
H.~{Desai}, M.~{Kantarcioglu}, and L.~{Kagal}, ``A hybrid blockchain
  architecture for privacy-enabled and accountable auctions,'' in \emph{2019
  IEEE International Conference on Blockchain (Blockchain)}, 2019, pp. 34--43.

\bibitem{8946175}
S.~{Zhu}, H.~{Hu}, Y.~{Li}, and W.~{Li}, ``Hybrid blockchain design for privacy
  preserving crowdsourcing platform,'' in \emph{2019 IEEE International
  Conference on Blockchain (Blockchain)}, 2019, pp. 26--33.

\bibitem{8946171}
T.~{Halevi}, F.~{Benhamouda}, A.~D. {Caro}, S.~{Halevi}, C.~{Jutla},
  Y.~{Manevich}, and Q.~{Zhang}, ``Initial public offering (ipo) on
  permissioned blockchain using secure multiparty computation,'' in \emph{2019
  IEEE International Conference on Blockchain (Blockchain)}, 2019, pp. 91--98.

\bibitem{8946149}
H.~{Guo}, W.~{Li}, M.~{Nejad}, and C.~{Shen}, ``Access control for electronic
  health records with hybrid blockchain-edge architecture,'' in \emph{2019 IEEE
  International Conference on Blockchain (Blockchain)}, 2019, pp. 44--51.

\bibitem{9169395}
H.~{Guo}, W.~{Li}, E.~{Meamari}, C.~C. {Shen}, and M.~{Nejad},
  ``Attribute-based multi-signature and encryption for ehr management: A
  blockchain-based solution,'' in \emph{2020 IEEE International Conference on
  Blockchain and Cryptocurrency (ICBC)}, 2020, pp. 1--5.

\bibitem{8946226}
T.~H. {Kim} and J.~{Lampkins}, ``Ssp: Self-sovereign privacy for internet of
  things using blockchain and mpc,'' in \emph{2019 IEEE International
  Conference on Blockchain (Blockchain)}, 2019, pp. 411--418.

\bibitem{li2019blockchain}
W.~Li, M.~Nejad, and R.~Zhang, ``A blockchain-based architecture for traffic
  signal control systems,'' in \emph{2019 IEEE International Congress on
  Internet of Things (ICIOT)}.\hskip 1em plus 0.5em minus 0.4em\relax IEEE,
  2019, pp. 33--40.

\bibitem{li2020privacy}
W.~Li, H.~Guo, M.~Nejad, and C.-C. Shen, ``Privacy-preserving traffic
  management: A blockchain and zero-knowledge proof inspired approach,''
  \emph{IEEE Access}, vol.~8, pp. 181\,733--181\,743, 2020.

\bibitem{9217515}
H.~{Guo}, W.~{Li}, M.~{Nejad}, and C.~C. {Shen}, ``Proof-of-event recording
  system for autonomous vehicles: A blockchain-based solution,'' \emph{IEEE
  Access}, vol.~8, pp. 182\,776--182\,786, 2020.

\bibitem{baza2019b}
M.~{Baza}, N.~{Lasla}, M.~{Mahmoud}, G.~{Srivastava}, and M.~{Abdallah},
  ``B-ride: Ride sharing with privacy-preservation, trust and fair payment atop
  public blockchain,'' \emph{IEEE Transactions on Network Science and
  Engineering}, pp. 1--1, 2019.

\bibitem{etherwhitepaper}
F.~Vogelsteller, V.~Buterin \emph{et~al.}, ``Ethereum whitepaper,''
  \emph{Ethereum Foundation}, 2014.

\bibitem{kanza2018cryptotransport}
\BIBentryALTinterwordspacing
Y.~Kanza and E.~Safra, ``Cryptotransport: Blockchain-powered ride hailing while
  preserving privacy, pseudonymity and trust,'' in \emph{Proceedings of the
  26th ACM SIGSPATIAL International Conference on Advances in Geographic
  Information Systems}, ser. SIGSPATIAL ’18.\hskip 1em plus 0.5em minus
  0.4em\relax New York, NY, USA: Association for Computing Machinery, 2018, p.
  540–543. [Online]. Available: \url{https://doi.org/10.1145/3274895.3274986}
\BIBentrySTDinterwordspacing

\bibitem{baza2020light}
M.~{Baza}, M.~{Mahmoud}, G.~{Srivastava}, W.~{Alasmary}, and M.~{Younis}, ``A
  light blockchain-powered privacy-preserving organization scheme for ride
  sharing services,'' in \emph{2020 IEEE 91st Vehicular Technology Conference
  (VTC2020-Spring)}, 2020, pp. 1--6.

\bibitem{semenko2019distributed}
Y.~Semenko and D.~Saucez, ``Distributed privacy preserving platform for
  ridesharing services,'' in \emph{Security, Privacy, and Anonymity in
  Computation, Communication, and Storage}, G.~Wang, J.~Feng, M.~Z.~A. Bhuiyan,
  and R.~Lu, Eds.\hskip 1em plus 0.5em minus 0.4em\relax Springer International
  Publishing, 2019, pp. 1--14.

\end{thebibliography}
\end{document}